\documentclass{article}
\usepackage{epsfig}
\usepackage{graphicx}

\newcommand{\etal}{et~al.}

\title{VLBA Scientific Memorandum  31: \\
  ASTROMETRIC CALIBRATION OF mm-VLBI USING SOURCE FREQUENCY PHASE
  REFERENCED OBSERVATIONS}

\author{Richard Dodson and  Mar\'{\i}a  Rioja\\
University of Western Australia, UWA, Australia \\
Observatorio Astron\'omico Nacional, Espa\~na}

\begin{document}
\maketitle
\pagebreak

\begin{center}
\large
{\bf ABSTRACT} \\
\end{center}

\normalsize
\noindent
In this document we layout a new method to achieve ``bona fide'' high
precision Very Long Baseline Interferometry (VLBI) astrometric
measurements of frequency-dependent positions of celestial sources
(even) in the high (mm-wavelength) frequency range, where conventional phase
referencing techniques fail.  Our method, dubbed {\sc Source/Frequency
  Phase Referencing} ({\sc sfpr}) combines {\it fast frequency switching}
(or dual-frequency observations) with the {\it source switching} of conventional
phase referencing techniques.  The former is used to calibrate the
dominant highly unpredictable rapid atmospheric fluctuations, which
arise from variations of the water vapor content in the troposphere,
and ultimately limit the application of conventional phase referencing 
techniques;  the latter 
compensates
the slower time scale remaining ionospheric/instrumental,
non-negligible, phase variations.\\

\noindent
For cm-VLBI, the {\sc sfpr} method is equivalent to conventional phase 
referencing applied to the measurement of frequency-dependent 
source positions changes (``core-shifts'').
For mm-VLBI, the {\sc sfpr} method stands as the only approach which
will provide astrometry. A successful demonstration of the application
of this new astrometric analysis technique to the highest frequency VLBA
observations, at 86 GHz, is presented here.  Our previous comparative
astrometric analysis of cm-VLBI observations, presented elsewhere,
produced equivalent results using both methods.

In this memo we layout the scope and basis of our new method, along with a 
description of the strategy 
(Sections 1 and 2), and a demonstration of successful application to the
analysis of VLBA experiment BD119 (Section 3). Finally, in Section 4,
we report on the results from a series of 1-hour long VLBA experiments,
BD123A, B and C, aimed at testing the robustness of the method under a
range of weather conditions.

\section{Introduction}

One of the complications in VLBI, over connected array interferometers, arises
from the completely unrelated atmospheric conditions 
that the wavefronts propagate through before reaching the 
widely separated antennae.
%
Self calibration procedures, which are the standard VLBI analysis
technique for imaging radio sources, rely on closure relations to
remove the station dependent complex gain factors that characterize
the phase errors at each antenna.  A direct detection of the source
with good signal to noise ratio is required within every segment of
the coherence integration time-interval. This time interval is set by
the stability of the instrument and, dominantly, the atmospheric
turbulence.  An important consequence of the use of phase
closure is that information on the absolute position of the source is lost, preventing the measurement of astrometric quantities.\\

%
%
The application of phase-referencing  techniques, to the
analysis of interleaving observations of the program source and a
nearby calibrator, preserves the information on the
angular separation on the sky
and provides high precision relative-astrometry (Alef 1988, Beasley \& 
Conway, 1995).
%
At the observations, the scans on the scientifically interesting
source, the target, are interleaved (within the coherence integration
time) with observations of a calibrator source, the reference. The
antenna-based corrections derived from the self-calibration analysis
of the reference source observations are transferred for the
calibration of the target source. Next, the target dataset is Fourier
transformed without any further calibration to yield a phase
referenced map of the target source, where the position offset of the
peak from the center provides a precise measurement of the relative
separation between both sources. The propagation of astrometric errors
in the phase referencing analysis is strongly dependent on the angular
separation between the target and reference sources, and range between
the micro-arcsec and tenths of milli-arcsec accuracy. This phase
referencing technique, from now on referred as ``conventional phase
referencing'', is well established and has been used to provide high
precision astrometric measurements of (relative) source positions in
cm-VLBI observations.

%
It would be highly desirable to extend this capability to the mm-VLBI
regime, yet at the highest frequencies the observations
are sensitivity limited:
the instruments are less efficient, the sources are intrinsically
weaker, and the phase coherence integration times are severely 
constrained by the rapid atmospheric phase fluctuations due to the
variations (spatial and temporal) of the water vapor content in the 
troposphere.
In particular, the coherence time is too short to allow an antenna to
switch its pointing direction between pairs of sources, in all but the
most exceptional cases (Porcas \& Rioja, 2002), within that time
range. The lack of suitable reference sources in mm-VLBI makes it
almost impossible to apply ``conventional phase referencing'' techniques in
the high frequency (i.e. significantly above 43\,GHz) domain. \\

Therefore it would be hugely beneficial if the calibration could be
performed at a lower, easier, frequency and used for data collected at
a higher frequency. That is, to transfer the calibration terms (for
phase/delay/rate VLBI observables) derived at a different frequency
rather than at a different source as in ``conventional
phase-referencing''. It should be noted that frequency switching can
be performed much faster than source switching at the VLBA.  Moreover
the duty-cycle is now determined by the coherence time at the lower
frequency. The low frequency phases provide `connection' but of
course can not correct for variations faster than the duty cycle. This
requires co-temporal dual frequency observations as provided by the
next generation of VLBI antennae and arrays which are able to co-observe
at different frequency bands, e.g. the Yebes 40m antenna and the Korean
VLBI Network.\\



The feasibility of multi-frequency observations to correct the
non-dispersive tropospheric phase fluctuations in the high frequency
regime has been studied for some time.  It relies on the fact that
such fluctuations will be linearly proportional to the observing
frequency, and hence it should be possible to use a scaled version of
the calibration terms derived from the analysis of observations at a
lower frequency (where more and stronger sources are available, with longer
coherence integration times and better antenna performance), to
calibrate higher frequency observations. It is a kind of phase
referencing, between observations at two frequencies, that we call
``frequency phase transfer'' ({\sc FPT}).
Among the earliest references we found are ``Phase compensation
experiments with the paired antennas method 2. Millimeter-wave fringe
correction using centimeter-wave reference'' (Asaki, \etal\ 1998) with
the Nobeyama millimeter array (NMA), 
and ``Tropospheric Phase Calibration in
Millimeter Interferometry'' (Carilli \& Holdaway, 1999) for
application with the Very Large Array (VLA).
%
In ``VLBI observations of weak sources using fast frequency
switching'', Middelberg \etal\ (2005) applied this frequency phase
transfer technique to mm-VLBI observations. They achieved a
significant increase in coherence time, resulting from the compensation
of the rapid tropospheric fluctuations, but failed to recover the
astrometry, due to the remaining residual dispersive terms.
%

Our proposed {\sc Source/Frequency phase referencing} method endows
this approach with astrometric capability for measuring frequency
dependent source positions (``core-shifts'') by adding a strategy to
estimate the ionospheric (and other) contributions. In ``Measurement
of core-shifts with astrometric multi-frequency calibration'' (Rioja
\etal\ 2005) we applied it to measure the ``core-shift'' of quasar
1038+528 A between S and X-bands (8.3/2.2 GHz), and validate the
results by comparison with those from standard phase
referencing techniques at cm-VLBI, where both methods are equivalent. \\
Here we present a demonstration of successful application of the {\sc
  sfpr} method to astrometric mm-VLBI, a much more challenging
frequency regime where conventional phase referencing fails.  Also,
the basis of the method, and details on the scheduling and
data analysis are described. \\
This method opens a new horizon with targets and fields suitable for
high precision astrometric studies with VLBI, especially at high
frequencies where severe limitations imposed by the rapid fluctuations
in the troposphere prevent the use of conventional phase referencing
techniques. In addition this method can be applied to Space VLBI,
where accurate orbit determination is a significant issue. This method
results in perfect correction of frequency independent errors, such as
those arising from the uncertainty in the reconstruction of the
satellite orbit. The application to the space mm-VLBI mission VSOP-2
is described in detail in Rioja \& Dodson (2009).




\section{The Basis of the new astrometric method}

This section outlines the basis of an astrometric 
method aimed at measuring the frequency dependent core position shift
(``core-shift'' hereafter) in radio sources in the high frequencies regime.
The novel {\sc sfpr} approach consists of two calibration
steps:  \\

\begin{itemize}
\item Dual-frequency observations to calibrate the rapid
  non-dispersive atmospheric phase fluctuations in VLBI observables at
  the high frequency regime, arising from inhomogeneities in the water
  vapor content in the troposphere
and \\

\item  Dual-source observations 
  to compensate for the remaining dispersive slower varying
  contributions to the observed phases.\\

\end{itemize}

The first step results in increased coherence times at the higher frequency, 
however, an extra step of
calibration which involves observations of a second source 
is needed to preserve the astrometry. 
This is the essence of the {\sc SFPR} technique.

We include here a description of the procedure using conventional
formulae for VLBI - and assume that the data reduction is done using
AIPS. Dodson \& Rioja (2008) contains more prescriptive details.  Its
application involves observations with \underline{fast frequency
  switching} between the two frequencies of interest ($high$ and
$low$, shown as superscripts in the formulae, for the higher and lower
observed frequencies), and \underline{slow source switching} between
the target and a nearby source ($A$ and $B$, shown as subscripts in
the formulae).
Following standard nomenclature, the residual phase (that is
after a\,priori estimated values for the various contributing
terms have been removed and the signal integrated in the correlator)
values for observations of the target source ($A$) at the lower frequency
($low$) with a given baseline, $\phi^{low}_{A}$, are shown as a 
sum of contributions: \\

$
\phi^{low}_{A} = \phi^{low}_{A,geo} +\phi^{low}_{A,tro} +\phi^{low}_{A,ion}
+\phi^{low}_{A,inst} + \phi^{low}_{A,str} + 2\pi n^{low}_{A}
,\,\,\,\,\,$ with $n^{low}_{A}$ integer \\

\noindent
where $\phi^{low}_{A,geo}, \phi^{low}_{A,tro}, 
\phi^{low}_{A,ion}$ and $\phi^{low}_{A,inst}$ are, respectively, 
contributions to the
residual phase from geometric, propagation medium -- troposphere and
ionosphere -- and instrumental errors, and $\phi^{low}_{A,str}$ is the
radio structure term (the visibility phase), referenced to the point for
which $\phi^{low}_{A,geo}$ has been computed, which is non-zero 
for non-symmetric sources; $2\pi n$  stands for the modulo $2 \pi$ phase
ambiguity term.

The application of ``self-calibration'' techniques produces an
image of the source, which allows one to disentangle the effect of the 
visibility phase from the rest of contributions, and produce a set of 
antenna-based terms, $\phi^{low}_{A,self-cal}$, that account for the
errors mentioned above. 
 




These terms are scaled by the frequency ratio $R$, after
interpolation to the observing times of the higher frequency scan times
observations, $\tilde \phi^{low}_{A,self-cal}$, and used to calibrate the 
higher frequency observations. \\

The resultant frequency-referenced residual phases at the higher
frequency $\phi_A^{FPT}$ are:

$\hspace*{3.5cm} \phi_A^{FPT} = 
\phi_A^{high} - R\,.\,\tilde \phi_{A,self-cal}^{low} =\\
= \phi_{A,str}^{high} + (\phi_{A,geo}^{high} - R\,.\,\tilde \phi_{A,geo}^{low}) 
+ (\phi_{A,tro}^{high} - R\,.\,\tilde \phi_{A,tro}^{low}) 
+ (\phi_{A,ion}^{high} - R\,.\,\tilde \phi_{A,ion}^{low}) + \\
\hspace*{2.5cm} + (\phi_{A,inst}^{high} - R\,.\,\tilde \phi_{A,inst}^{low}) 
+ 2\pi (n^{high}_A-R\,.\,n^{low}_A) \hspace*{2cm} (1)$  \\ 


where $FPT$ stands for ``Frequency Phase Transfer'' and $\tilde \phi$
stands for the interpolated $low$-frequency self-calibration solutions
to the $high$-frequency scan times.  This calibration strategy results
in perfect cancellation (as long as the interpolation of $\tilde \phi$
is a good approximation to $\phi$) of the non-dispersive rapid
tropospheric phase fluctuation terms, since:

\begin{center}
$\phi_{A,tro}^{high} - R\,.\,\tilde \phi_{A,tro}^{low} = 0 \, $ \\
\end{center}

but not for the dispersive ones, which do not scale linearly with
frequency, and hence there are remaining ionospheric and instrumental terms: 

\begin{center}
$\phi_{A,ion}^{high} - R\,.\,\tilde \phi_{A,ion}^{low} = 
({1 \over R}-R)\, \tilde \phi_{A,ion}^{low}$ \\
$\phi_{A,inst}^{high} - R\,.\,\tilde \phi_{A,inst}^{low} \neq 0$ \\
\end{center}

Notice that while antenna and source coordinates errors,
given the non-dispersive nature of geometric terms, cancel out in
this calibration procedure, a frequency dependent source position shift
$\bar{\theta}_{A}$ would remain, since:

\begin{center}
$\phi^{high}_{A,geo}  - R\,.\,\tilde\phi^{low}_{A,geo}  = 
2\pi \,\vec{D_{\lambda}}\,.\,\vec{\theta_{A}}$\\
\end{center}

\noindent
where $\vec{D_{\lambda}}$ is the baseline vector, in units of
the higher wavelengths, and $\vec{\theta_{A}}$ stands for the ``core-shift''.

An integer frequency ratio $R$ will keep the phase ambiguity term in the
phase equations as an integer number of $2 \pi$, and avoid phase 
connection problems. We strongly advice to use the observations at a
given frequency ($low$) to calibrate the harmonic frequencies ($high$).
For simplicity we will omit the $2 \pi$ ambiguity term in the coming equations. 
When $n$ is zero $R$ does not need to be integer, see Rioja \etal\ (2005). 

\noindent
Replacing the relations above in equation (1), the expression for the
``frequency transferred'' residual phases for the observations of
source $A$ at the higher frequency becomes: \\


$\phi_A^{FPT} = 
 \phi_{A,str}^{high} + 
2 \pi \, \vec{D_{\lambda}}\,.\, \vec{\theta}_{A} 
+ ({1 \over R}-R) \, \tilde \phi_{A,ion}^{low}
+ (\phi_{A,inst}^{high} - R \,.\, \tilde \phi_{A,inst}^{low}) 
 \hspace*{1cm}  (2)$ \\ 

\noindent
The rapid tropospheric fluctuations have been calibrated out, however
longer timescale contaminating ionospheric and instrumental terms
remain blended with the radio structure and astrometric ``core-shift''
signature, and prevent its direct extraction from the phases.
Previous applications of the dual-frequency calibration method used an
extra step of self-calibration to remove these, with the consequent
loss of the frequency-dependent position of the source
(the ``core-shift'') in the sky (as in Middelberg \etal\ 2005).

We propose a different scheme that removes the non-dispersive terms
while preserving the astrometric information.  It uses the
procedure of interleaving fast-frequency switching observations of the
program source $A$ with those of a calibrator which is nearby in
angle, $B$, in a very similar fashion as it is done for conventional
phase referencing.


\noindent

\noindent
The analysis of the $B$ dataset is done following the same procedure as
for $A$, and arrive to an equivalent expression to equation (2) for 
the {\sc fpt}-phases of $B$: \\


$\phi_{B}^{FPT} =
 \phi_{B,str}^{high} + 
2 \pi \, \vec{D_{\lambda}}\,.\,\vec{\theta_{B}}
+ ({1 \over R}-R) \, \tilde \phi_{B,ion}^{low}
+ (\phi_{B,inst}^{high} - R \,.\, \tilde \phi_{B,inst}^{low}) 
\hspace*{1cm} (3)$ \\ 

\noindent

A careful planning of the observations, namely alternating between two
sources that lie within the same ionospheric isoplanatic patch (whose
size is many degrees at mm-wavelengths) with a duty cycle that matches
the shortest ionospheric/instrumental time-scales (several minutes at
least), results in the remaining dispersive terms in equations (2) and
(3) being close to equal. That is:

\begin{center}
$\hspace*{1cm} ({1 \over R}-R) \, \tilde \phi_{A,ion}^{low} \approx ({1 \over
  R}-R) \, \tilde \phi_{B,ion}^{low} $ \\
$\hspace*{1.5cm} \phi_{A,inst}^{high} - R \,.\, \tilde \phi_{A,inst}^{low} \approx 
\phi_{B,inst}^{high} - R \,.\, \tilde \phi_{B,inst}^{low} $ \\
\end{center}

\noindent
Under these conditions the {\sc fpt}-phases of $B$ can be used to
calibrate the $A$ dataset, as in ``conventional phase referencing''.
That is, apply self-calibration techniques on the {\sc fpt}-phases
from the $B$ dataset (including removal of the structure contributions),
and transfer the estimated antenna-based corrections for the calibration
of the {\sc fpt}-phases of $A$, after interpolation to the
corresponding observing times, $\tilde \phi^{FPT}_{B,self-cal}$.




\noindent
The resultant {\sc Source/Frequency-referenced} residual phases for
the target source $A$, $\phi^{SFPR}_{A}$, are free of ionospheric/instrumental 
corruption while keeping the astrometric ``core-shift'' signature: \\

$\phi^{SFPR}_{A} = \phi^{FPT}_{A} - \tilde \phi^{FPT}_{B,self-cal}$ 
$= \phi_{A,str}^{high} + 2 \pi \vec{D_{\lambda}} \, . \, (\vec{\theta}_{A}-
\vec{\theta_{B}})$ \\  

where $\phi_{A,str}^{high}$ stands for the 
radio structure contribution of source $A$ at the high frequency,
and the terms $2\pi \, \bar{D}_{\lambda} \,.\, \bar{\theta}_{A}$ and 
$2\pi \, \bar{D}_{\lambda}\,.\, \bar{\theta}_{B}$ 
modulate each baseline with a $\sim 24$ hours period sinusoid  
whose amplitude depends on the ``core-shifts'' in $A$ and $B$, respectively
- and is equivalent to the functional dependence on the source pair angular
separation in ``conventional phase referencing''.\\

\noindent
Finally, the calibrated {\sc sfpr}-visibility phases from the target
source $A$, $\phi^{SFPR}_{A}$, are inverted to yield a synthesis image
of source $A$ at the $high$-band, where the offset from the center
corresponds to a bona-fide astrometric measurement of the combined
frequency dependent ``core shifts'' in sources $A$ and $B$ between the
$low$ and $high$-frequency bands.\\





\noindent
We have summarised the contributions to the residual phases and how
they are handled in our new strategy for carrying out astrometric {\sc
  Source/Frequency Phase Referenced} observations.  Because of the
large calibration overhead involved in {\sc sfpr} this method is only
recommended for mm-VLBI, where no other method would succeed.
Previous efforts to exploit the astrometric application of
multi-frequency techniques failed (even for mm-VLBI), because of what
is believed to be the ionospheric contribution. Whilst improved phase
stabilisation was achieved, and the deepest ever detection of VLBI
cores at 86-GHz were produced, astrometric results could not.
Our improved method compensates the remaining ionospheric and
instrumental contributions while preserving the astrometric signature
in the calibrated visibilities, and, of course, also increases the
coherence integration time of the observations at the higher
frequency.

We have not yet addressed the errors introduced by the interpolation
of the lower frequency phase to the times of the high frequency
observations (i.e. $\phi^{low}$ to $\tilde \phi^{high}$), nor the
constraints on the frequency switching duty cycle, both closely related to
the coherence at the time of the observations.
Using a typical value for the Allan standard deviation
($\sigma_{Allan}$) of atmospheric phase fluctuations equal to
$10^{-13}$ over 100 seconds (Thompson, Moran, Swenson 2001) and the
accumulated phase noise ($\Delta \phi_{a} = 2\pi \nu \sigma_{Allan}
\Delta {\rm t} $) one can estimate the coherence time ($\Delta {\rm
  t}) $ for a one radian change in phase. This results in typical
coherence times of about 70 and 40\,seconds, at 22 and 43\,GHz,
respectively.
The duty cycle for the frequency switching has to be less than this
coherence time for the lower frequency, if the conditions are
typical. However normally one would normally request `better than
typical' weather conditions for mm-VLBI which would improve these
limits.
%
The error due to the interpolation of the low frequency phases to the
time of the high frequency observations can be estimated from the
errors ($\sigma_{low,i}$ for scan $i$) before and after the high frequency scan
multiplied by the frequency ratio ($R$). If the duty cycle equals the
coherence time the low frequency observations before and after the
high frequency scan are independent, but (assumed to be) smoothly
varying. Therefore, taking the errors as equal for the bracketing observations,
the error on the high frequency observation is given by
$R\sigma_{low}$. If the duty cycle is much less than the coherence
time the errors will be $R\sigma_{low}/\sqrt{2}$ (as the measurements
are independent, but the observables are not) however this will
involve inefficient use of time in switching the frequencies. If the
duty cycle is greater than the coherence time the solutions can not be
connected with a linear extrapolation and the calibration will fail.

    One can then deduce the minimum SNR required in the low frequency
    scans to ensure that the estimated phase corrections for the high
    frequency observations are meaningful. Assuming that $\sigma_{low}$ is
    given by the rms thermal phase noise formula ($\sigma_{low} \sim
    1/$SNR) and setting an upper limit of $30^o$
    for the high frequency phase correction estimates, ones imposes SNRs in
    the low frequency scans equal to 6 and 11, for observations at
    frequency pairs 22/43 GHz and 22/86GHz, respectively. Note also that if one
    was choosing between using 22- or 43-GHz as the calibrator for an
    86-GHz target one needs to balance the halving of the frequency
    ratio against the approximate three times higher SEFD for 43-GHz
    observations.



    As an aside, the Korean VLBI Network (KVN), the world's first
    dedicated mm-VLBI array, will be able to observe 4 bands simultaneously
    (22/43/86/129 GHz). This will remove the need for frequency
    switching, tripling the observing time in a typical implementation
    thereby increasing the SNR, and furthermore remove the need for
    interpolation hence reducing the accumulated errors. 

\section{Demonstrations of the Method and Results}

On 18 February 2007 we carried out 7 hours of VLBA observations of two
pairs of continuum sources (1308+326 \& 1308+328 $14^\prime$ apart,
and 3C273 \& 3C274 $10^o$ apart), using \underline{fast frequency switching}
between 43- and 86-GHz scans on each source, and \underline{slow antenna
switching} between the sources, for each pair.  Each antenna recorded
eight 16-MHz IF channels, using 2-bit Nyquist sampling, which resulted
in a data rate of 512 Mbps.
%
The total duration of the observations was divided in
$\sim$1.5-hour long blocks allocated to alternate observations of 
the two source pairs.

%
The analysis was done mostly using AIPS. See Dodson \& Rioja (2008) for
more details on the tasks and considerations.
Firstly we followed the general VLBI calibration procedures,
using the scans on the primary calibrator (3C273), and applied it
to the total duration of the observations.
%
Next, for each pair, we applied self-calibration analysis procedures
to the observations of the two sources at 43\,GHz (the ``lower''
frequency), scaled the resulting phase terms by a factor of 2 (using
an external perl script), and applied these to the same source's
observations at 86\,GHz (the ``higher'' frequency).
%
Then, we ran self-calibration procedures on the strongest source of
each pair, 1308+326 and 3C273, respectively, at 86\,GHz. These
solutions were then transferred for calibration to the observations of
the other source's pair, 1308+328 and 3C274, respectively, at the higher
frequency, which were finally imaged without further calibration.
%
The result of the analysis is, for each pair, is a {\sc sfpr} map which
contains the brightness distribution of the target source at 86\,GHz,
and where the offset of the peak of flux from the centre is
astrometrically significant, and corresponds to the combined relative
``core-shift'' between 43 and 86\,GHz, of both sources.

An additional complication that we have not mentioned above is related
to the source radio structure effects; this is relevant when the
sources are extended, as it is the case for the 3C pair.  For this
case both the lower frequency FRING phase solutions and the CALIB
solutions (based on the best hybrid image) were doubled and applied
for the calibration of the higher frequency observation. For compact
sources, as it is the case for the 1308+326/8 pair, the structure
contribution
is negligible.\\


Figure 1 shows the {\sc sfpr} image of 1308+328, at 86 GHz. This map
was made following the method described above, using the 43-GHz
observations of the source and further corrections derived from
1308+326, $14^\prime$ away.  The flux recovery in our image, defined
as the ratio between the brightness peaks in the {\sc sfpr}-map to the
hybrid map, is 60\%. For comparison, the only ``conventional phase
referenced'' map which has been done at 86-GHz (Porcas \& Rioja 2002),
on this same pair of sources, resulted in a flux recovery of only
20\%. Previous multi frequency VLBI observations of this pair of
sources (Rioja \etal\ 1996) are compatible with a zero ``core-shift'',
as found in our analysis.

Figure 2 shows the {\sc sfpr} image of 3C274 (M87) at 86\,GHz. The
larger angular separation between the sources in the 3C pair, $10^o$
apart, compared to the $14^\prime$ for the 1308+32 pair, makes this
case a more challenging test. Still, the flux recovery in the {\sc
  sfpr} image in Figure 2 is 60\%.  The peak of brightness does show
an offset from the centre of the map equal to 70$\mu$sec.  In the
absence of any other observations to compare our results with, we note
here that the predicted ``core-shifts'' for 3C273
is 65 $\mu$sec, and zero for 3C274 (Lobanov, 1998 for 3C273 and
personal comms. for 3C274), so it is possible that we again have a
correct solution. As the theoretical predictions are not, at best, an
exact science, it would be wise instead to take the measured
``core-shift'' in the map as an order of magnitude estimate of the
reliability of the method. That is, we give an upper bound of 0.1
milli-arcsec to the astrometric accuracy produced by this strategy,
pending further investigation.

\section{Robustness of the Method vs. Weather}
This method has also been shown to work well, in terms of astrometric
recovery, in observations made without weather constraints.  We
summarize here the results found from the analysis of a series of
1-hour long VLBA test observations (Exp. codes BD123A/B/C) of the pair
of sources 1308+326/8, with the VLBA, at 43 and 86 GHz.
We followed the procedure described in section 3 for the analysis of
the observations to produce {\sc sfpr}ed maps of 1308+328 at 86GHz, 
and also applied self-calibration procedures to produce hybrid
maps, using AIPS.\\
The datasets were inspected and some baselines were flagged out based on
lack of detections of the calibrator source. \\
To assess the quality of the results in the three sessions we used the 
ratio of the brightness peak values in the {\sc sfpr}ed to the hybrid maps,
which we refer to as ``flux recovery'' hereafter.
Calibration errors, arising from imperfect phase compensations in the
analysis using a reference source/frequency, 
are responsible for the decrease, and biased positional offsets, 
of the peak flux
in the {\sc sfpr}ed map.


%
Figure~3 shows the {\sc sfpr}ed maps of 1308+328 at 86-GHz obtained
from the analysis of these observations.  The map corresponding to
the analysis of BD123B ({\it center}) shows 88\% ``flux recovery''
with good solutions for all antennae in the dataset; for BD123C ({\it
  right}) the flux recovery is 60\% with failed solutions for the NL and
LA antennae; and for BD123A ({\it left}) the flux recovery is only
23\%, with failed solutions the LA and PT antennae.
The low ``flux recovery'' in the map from BD123A rises 
doubts about the success of the technique in that case.
However the location of the peak flux in the three maps, which carries
the astrometric information, does not significantly shift from the centre, with
differences in positions $<20\,\mu$as.
%

The weather conditions are expected to have an impact on the quality
of the {\sc sfpr} analysis results, as happens in conventional phase
referencing.  It would be very useful to have a threshold criterion
for successful {\sc sfpr} with respect to the weather conditions,
especially at the highest frequencies. We could not address this
question in this series of three test experiments since the weather
``predictions'' were not kept after the observations. We could not
find either any outstanding correlation between the ground
meteorological data measured during the observations, or $T_{sys}$
values, and the image quality for the three experiments. Instead, we
have attempted to characterize the weather, a posteriori, with phase
coherence measurements using the observations themselves.  We have
used a four minutes long scan on the calibrator source 3C273 at 43
GHz, after preliminary calibration using a 2-minutes solution interval
in FRING, for the three experiments.
For each experiment the scans were segmented at different intervals
(from 10 sec to 150 sec long, with 10 sec steps) and averaged (in
scalar and vector fashion) to determine the visibility amplitudes.
Figure~4 shows baseline phase coherence plots for the three
experiments: the amplitude ratio between the vector average and scalar
average in y-axis, against integration time in x-axis, for all
baselines to the reference antenna in each of the three experiments.
The plots only include baselines successfully detected in each
session, using different colours for each antenna (in baselines with
the reference antenna); the MK antenna (pink) shows the poorest
coherence and the lowest ($\sim30$) evelations in all the 3
experiments.

These experiments are so short (1-hour long) that 
this sample serves as a good indication of the conditions through-out
the experiment.
The plot for BD123B ({\it centre}) shows the best performance -
better than can be resolved in the 2 min data-span.  The coherence is
certainly more than several minutes at 43\,GHz, and corresponds to
the session with best image recovery.
%
For BD123C ({\it right}) the plot shows that severe coherence losses
occur quite quickly, but then they reach a plateau and stabilize. The
flux recovery in the map is 60\%.  Therefore we can say that the
2-minute solution interval is acceptable in this case.
In experiment BD123A, the coherence shows a steady decrease 
across the span of the data.  This is in agreement with the poor flux
recovery, of only 23\%, which would normally be described as a failure in the
phase referencing process, although we repeat that the position of the
peak of emission in the map is in agreement, to within 20\,$\mu$as,
with those from B and C datasets.

Based on Figures 3 \& 4, a tentative classification of the weather
conditions during the three 1-hour long observations, is: ``good'' for
BD123B, ``acceptable'' for BD123B and ``bad'' for BD123A. The
coherence times probe the weather conditions relevant for
mm-wavelength observations, such as the content of water vapor in the
troposphere, unobtainable from ground only meteorological
measurements.  We conclude that a suitable frequency duty cycle for
obtaining a good {\sc sfpr} map should be selected based on the
observed frequencies and particular weather conditions during the
observations; certainly it should be well under the coherence time at
the reference (lower) frequency at the epoch of the observations.
We are unable to give further guidance without further observations.

\section{Summary}

We have proposed and demonstrated a new method of astrometric VLBI
calibration, suitable for mm-VLBI. It uses dual frequency observations
to removed non-dispersive contributions.
The additional step required to remove the ionospheric, and all other
slowly varying dispersive terms, is done by including another source
to cross calibrate with. Because the ionospheric patch size is very
large at mm wavelengths one can use calibrators that lie a
considerable distance from the source.  We have presented the results
from two pairs of sources, one only $14^\prime$ apart and the other
10$^o$ apart.
A single pair is sufficient to demonstrate the method, however the
astrometric solution (the offset from expected position) contains the
contribution from both sources (as happens in standard
phase-referencing as well). This problem, however, fulfills the
closure condition, so three or more sources can be used to form a
closure triangle and separate the contributions from each individual
source.

\section*{Acknowledgments}
We wish to express our gratitude to Craig Walker, Ed Fomalont, Asaki
Yoshiharu and Richard Porcas for their help in the development of the
technique and in proof-reading this manuscript.

\paragraph{Acknowledgments} we wish to note the essential help of the
EU Marie-Curie International Incoming Fellowship
(MIF1-CT-2005-021873), the VLBA which in funded by the National
Science Foundation (of the USA) and the support of and advice from Ed
Fomalont, Richard Porcas and Vivek Dhawan.

\begin{center}
\begin{figure}[htb]
\epsfig{file=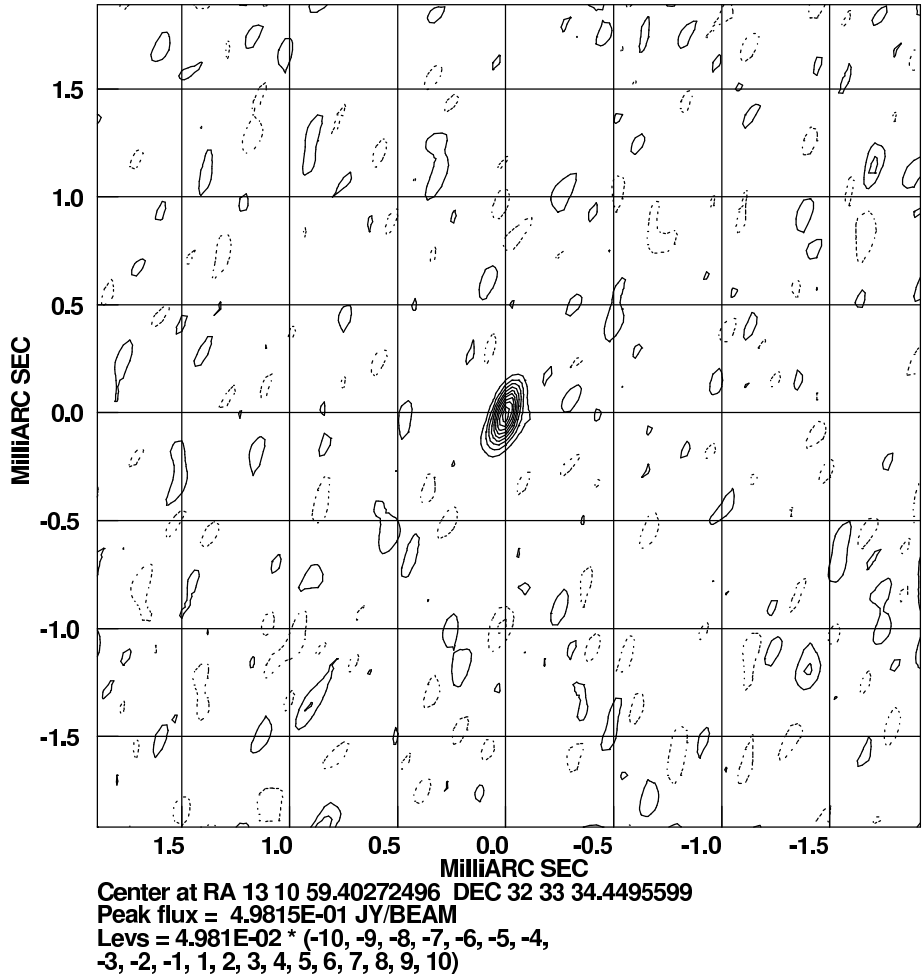,angle=0,width=\textwidth}
\caption{{\sc SFPR} map of 1308+328, at 86\,GHz, calibrated with the
  43\,GHz phases and further corrections from 1308+326 --
  $14^{\prime}$ away. The offset from the phase centre is zero
  to within the errors, as expected from other phase-referencing
  experiments. The flux recovery compared to the hybrid map is 60\%.}
\label{fig:f1}
\end{figure}

\begin{figure}[htb]
\epsfig{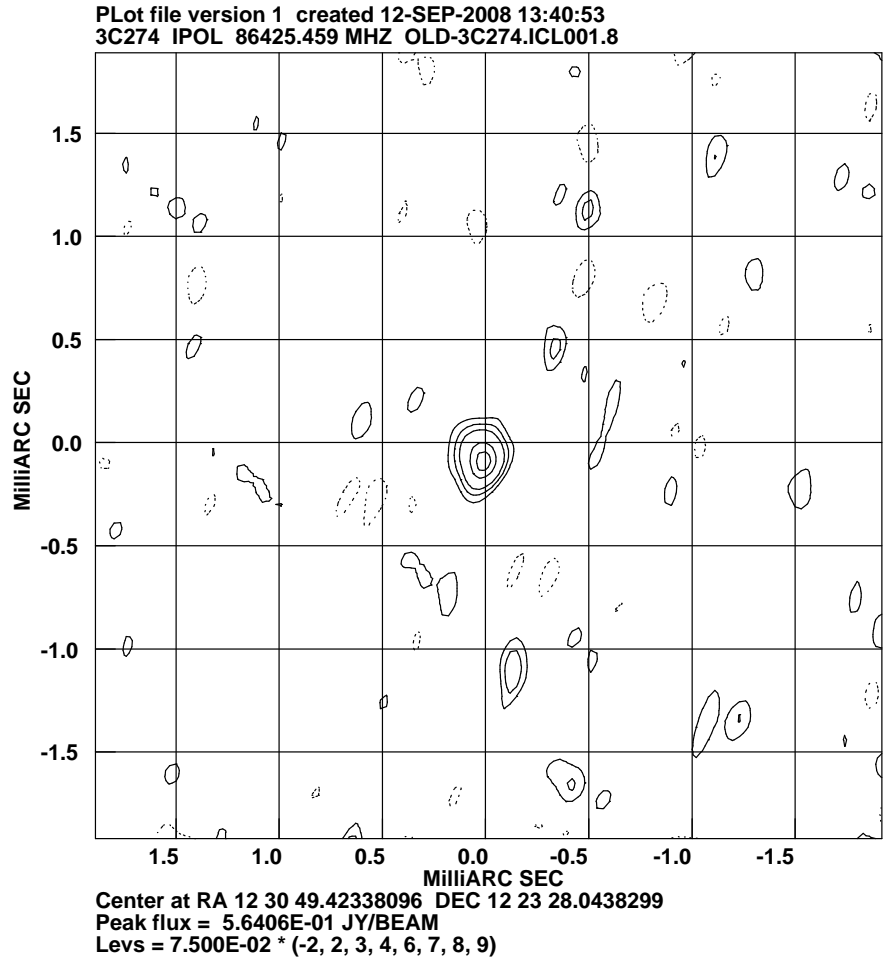}
\caption{{\sc SFPR} map of 3C274, at 86\,GHz, calibrated with the
  43\,GHz phases and further corrections from 3C273 -- $10^o$ away.
  The offset from the phase centre (70\,$\mu$sec) is close to that
  predicted on theoretical grounds. The flux recovery compared to the
  hybrid map is 60\%.}
\label{fig:f2}
\end{figure}


\begin{figure}[htb]
\epsfig{file=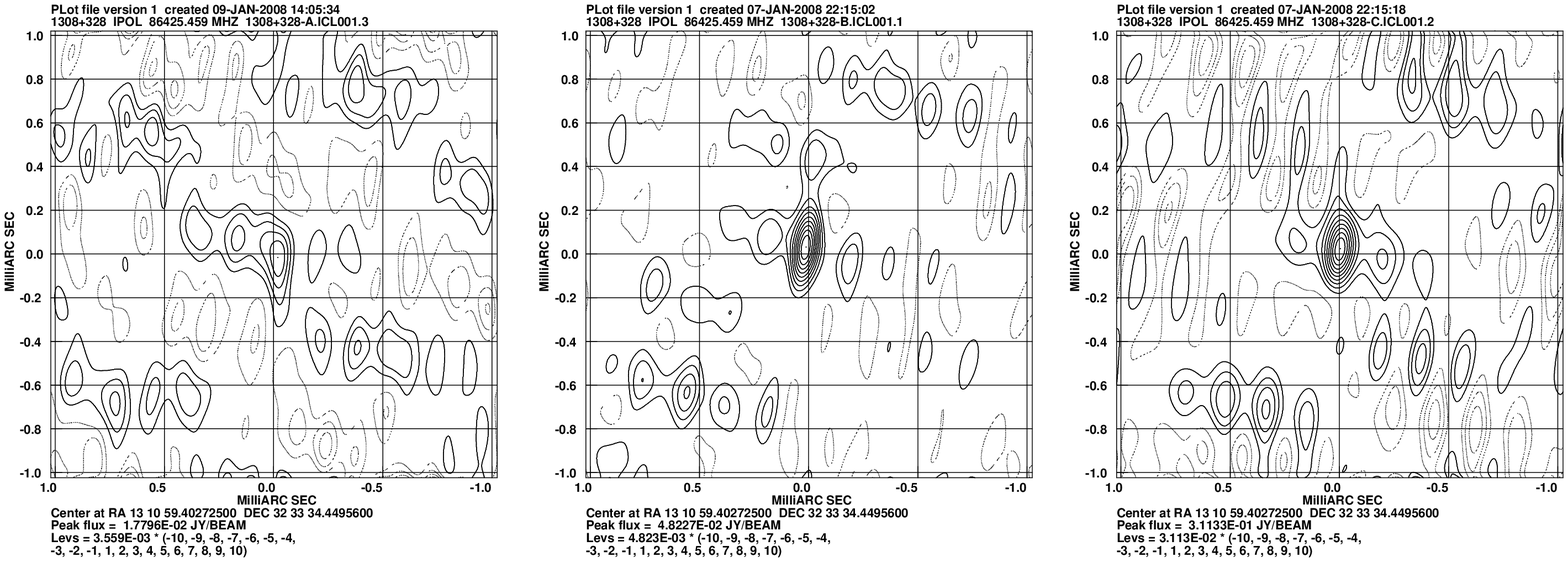,angle=0,width=\textwidth}
\caption{Montage of {\sc SFPR} maps of 1308+328, at 86\,GHz, from our
  1-hour long VLBA observations. The flux recovery compared to hybrid
  maps is 88\% for BD123B (centre), 60\% for BD123C (right) and 23\%
  for BD123A (left).  All experiments produce comparable astrometric
  results, to 20$\mu$arcsec.}
\end{figure}

\begin{figure}[htb]
\epsfig{file=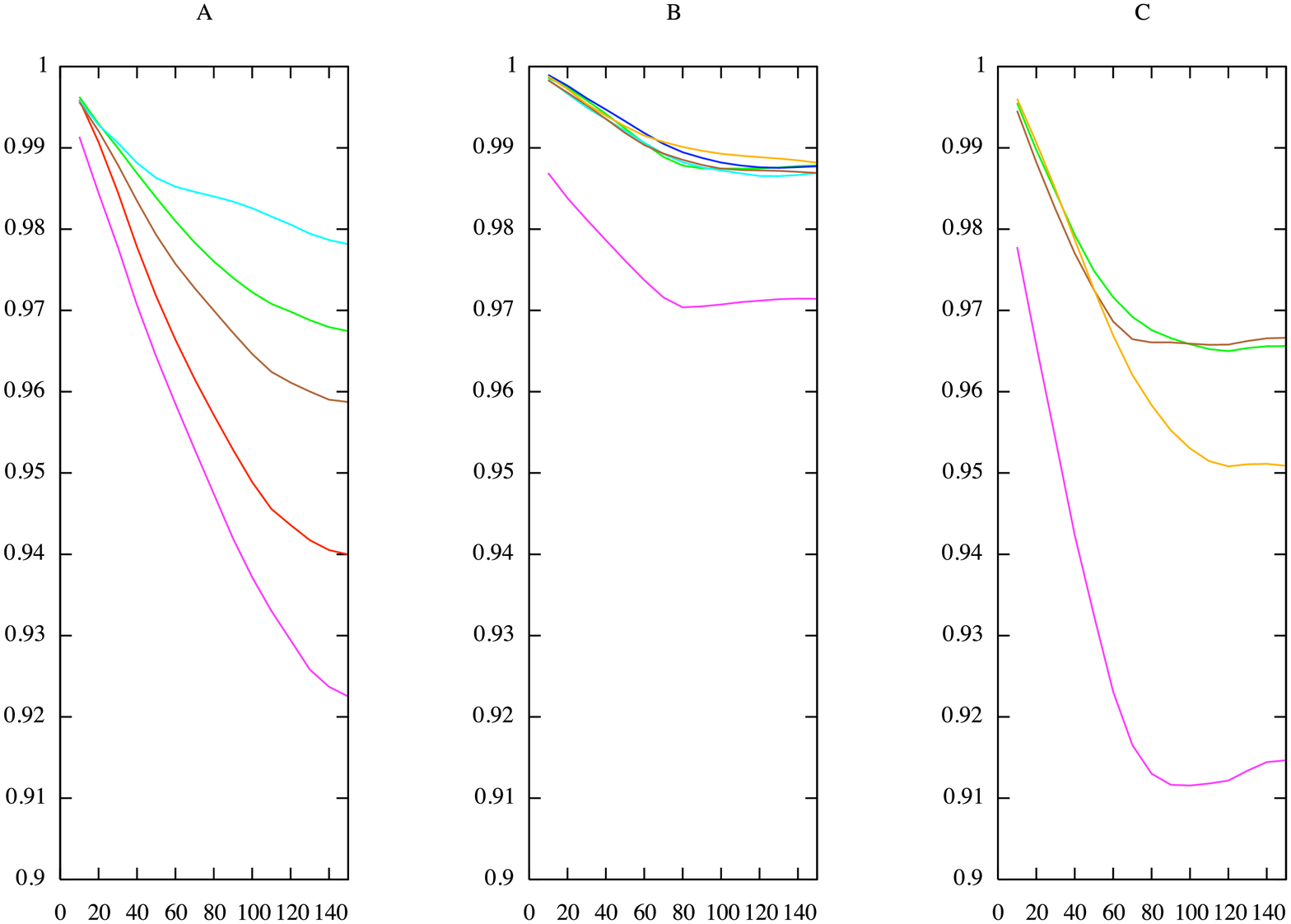,angle=0,width=\textwidth}
\caption{Coherence of the data for each of the three epochs of BD123.
  Plotted is coherence (vector average over scalar) against time in
  seconds on the primary calibrator 3C273. The different antennae are
  plotted in different colours.  The best weather (B, centre) produces
  the best flux recovery (88\%), the worst results (23\%) come from
  the worst weather (A, left), however the weather for the third epoch
  (C,right) is not much better however the flux recovered is 60\%.}
\label{fig:coh}
\end{figure}
\end{center}


\begin{thebibliography}{}
\bibitem{} Asaki, Y.,  \etal\ 1998.\ ``{Phase Compensation experiments
  with the paired antennas method 2. Millimeter-wave fringe correction
  using centimeter-wave reference.}'' \ Radio Science 33, 1297-1318.


\bibitem{} Carilli, C.~L., Holdaway, M.~A.\ 1999.\ ``{Tropospheric phase
  calibration in millimeter interferometry.}''\ Radio Science 34,
  817-840.

\bibitem{} Dodson, R.. Rioja, M., 2008, ``{On the astrometric
    calibration of mm-VLBI using dual frequency observations}'',
  IT-OAN-2008-3

\bibitem{1998A&A...330...79L} Lobanov, A.~P.\ 1998.\ ``{Ultracompact jets in active galactic nuclei.}''\ Astronomy and Astrophysics 330, 79-89. 

\bibitem{2005A&A...433..897M} Middelberg, E., Roy, A.~L., Walker, R.~C., Falcke, H.\ 2005.\ ``{VLBI observations of weak sources using fast frequency switching.}''\ Astronomy and Astrophysics 433, 897-909. 


\bibitem{2002evn..conf...65P} Porcas, R.~W., Rioja, 
M.~J.\ 2002.\ ``{VLBI phase-reference investigations at 86 GHz.}''\ Proceedings 
of the 6th EVN Symposium 65. 

\bibitem{2005astro.ph..5475R} Rioja, M.~J., Dodson, R., 
Porcas, R.~W., Suda, H., Colomer, F.\ 2005.\ ``{Measurement of core-shifts 
with astrometric multi-frequency calibration.}''\ ArXiv Astrophysics e-prints 
arXiv:astro-ph/0505475. 

\bibitem{}Rioja, M.~J., Porcas, R., Machalski. J.\ 1996.\ ``{EVN
  Phase-Referenced Observations of 1308+328 and 1308+326.}''\ Extragalactic
  Radio Sources, IAU Symp. 175, p. 122. 

\bibitem{TMS}Thompson, Moran, Swenson, 2001, ``{Interferometry and
    Sythesis in Radio Astronomy}'' (New York: John Wiley \& Sons)

\end{thebibliography}
\end{document}